\documentclass[aps,prl,twocolumn,groupedaddress,showpacs,showkeys]{revtex4}

\usepackage{graphicx}

\newcommand{\sqhz}{\ensuremath\sqrt{\mbox{Hz}}}
\newcommand{\rtHz}{\ensuremath\sqrt{\mbox{Hz}}}
\newcommand{\uVhz}{\ensuremath\mu\mathrm{V}/\sqrt{\mbox{Hz}}}

\begin{document}

\author{S. E. Pollack
\footnote{Present address: 
    Department of Physics and Astronomy and Rice Quantum Institute, 
    Rice University, Houston, TX  77251}
}
\email[]{skotep@skotep.com}
\author{S. Schlamminger}
\author{J. H. Gundlach}
\affiliation{Department of Physics, University of Washington, 
    Seattle, WA  98195-4290}

\date{\today}

\title{Temporal Extent
of Surface Potentials between Closely Spaced Metals}

\begin{abstract}
Variations in the electrostatic surface potential between the
proof mass and electrode housing in the space-based gravitational
wave mission LISA is one of the largest contributors of noise at
frequencies below a few mHz.
Torsion balances provide an ideal testbed for investigating
these effects in conditions emulative of LISA.
Our apparatus consists of a Au coated Cu plate brought
near a Au coated Si plate pendulum suspended from a thin W wire.  
We have measured a white noise level of $30\,\uVhz$ above
approximately 0.1\,mHz, rising at lower frequencies,
for the surface potential variations between these two closely spaced metals.
\end{abstract}


\maketitle


The low frequency sensitivity of the ESA/NASA 
gravitational wave mission LISA 
is limited by 
spurious accelerations of its enclosed proof masses \cite{PPA, lisaweb}.
Above 0.1\,mHz one of the largest contributors of
spurious accelerations are forces due to 
electrostatic patch field fluctuations \cite{merk}.  
Each of the proof masses (PM) in LISA are contained
within an electrode housing, together known as the gravitational
reference sensor (GRS).  The electrodes and the PM form a
collection of capacitors used in position determination and
proof mass actuation for the drag-free operation of each spacecraft.
Fluctuations in the electric potential between the PM
and each of the electrodes will lead to acceleration disturbances
of the PM.
The current LISA requirement on surface potential fluctuations,
based on a noise budget flowdown,
is that they do not exceed $50\,\uVhz$ above 0.1\,mHz \cite{LISA_Tech_Status}.
Previous measurements of surface potential variations,
using a Kelvin probe \cite{Norna} or a mock-up of the 
proposed LISA GRS which has finished engineering testing for
the LISA Pathfinder mission \cite{Carbone},
are believed to be measurement limited at 
$\sim 1\,$mV$/\rtHz$ above ~0.1\,mHz.

\begin{figure}
  \includegraphics[width=1.0\columnwidth,angle=0]{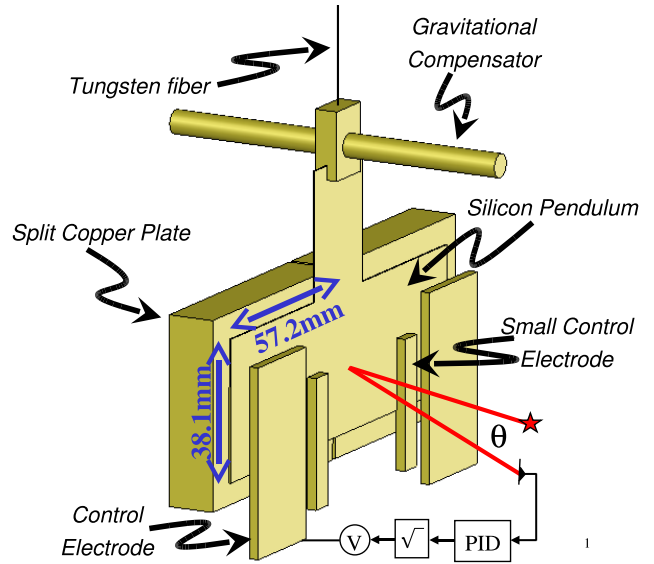}
\caption{
Schematic showing the Si pendulum suspended
from the W fiber with the split Cu plate and control electrodes.
The autocollimator for optical readout of the pendulum angle is shown
along with the feedback control loop mentioned in the text.
\label{fig:setup}}
\end{figure}

Our torsion balance apparatus has been specifically designed
to investigate parasitic voltage fluctuations and thermal gradient
related effects \cite{Stephan, lisa6}.
The pendulum is a mostly rectangular wafer
of Si suspended by a 53\,cm long, 13~$\mu$m diameter, W fiber.  
The Si was coated with an adhesion layer of $\sim20$\,nm TiW
and then $\sim225$\,nm layer of Au.
A movable Cu plate of slightly larger size is split into two halves, left and right.
It is coated with $\sim30$\,nm TiW then $\sim100$\,nm Au,
A schematic of our setup is shown in Figure~\ref{fig:setup}.
The separation between the pendulum and Cu plate can be adjusted
between 0 and 10\,mm to within $\approx 10\,\mu$m reproducibility.

Four Au coated control electrodes made of Al are mounted on the side
of the pendulum opposite that of the Cu plate, two smaller in size
and closer to the pendulum rotation axis.  The distance between
the control electrodes and the pendulum is $7\pm1$\,mm.  
Optical readout of the pendulum rotational angle is done by an autocollimator,
which is also located on this side of the pendulum.
The autocollimator combined with the control electrodes
provide a feedback mechanism for fixing the pendulum angular position.
The torque due to thermal and other forces on the 
pendulum can be measured through the control voltage
supplied to the control electrodes, typically the larger electrodes.
As detailed in \cite{Stephan}, running the system in 
the feedback mode does not increase the noise
or degrade the performance of our torsion balance.
The residual motion of our pendulum is at the level
of 2\,nrad/$\sqhz$, which is equivalent to a cold
damped temperature of about 0.3\,K.
Thermal noise at 297\,K is nearly 1\,$\mu$rad/$\sqhz$
at 1\,mHz.

The entire assembly is housed within a vacuum chamber with a
base pressure $\approx 10^{-5}$\,Pa.  
A polystyrene enclosure around the vacuum chamber 
assists in passive thermal control of the apparatus.
The chamber rests on a large Al plate which 
is supported on a block of concrete.

\begin{figure}
  \includegraphics[width=0.75\columnwidth,angle=90]{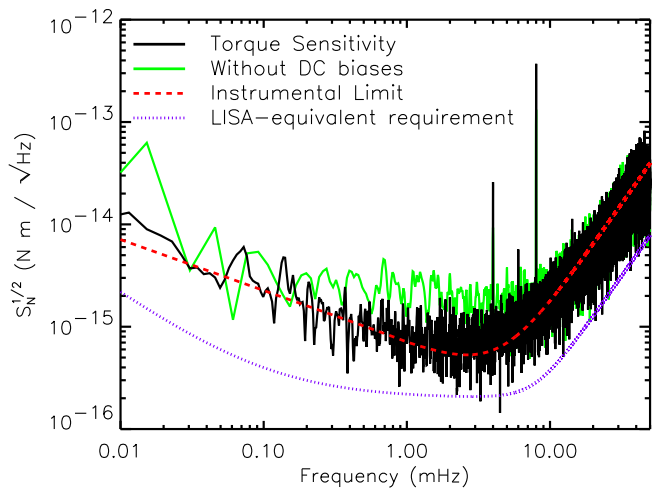}
\caption{
(color online) Torque sensitivity of our apparatus in feedback 
with a plate-pendulum separation of 1\,mm with (dark)
and without DC biases applied (green/grey).
Our instrumental limit (dashed) and the LISA-equivalent 
requirement (dotted) are also shown.
The quadrupole and hexadecapole gravitational
couplings, along with an off-axis coupling, 
of our external calibration source 
can be seen at 4, 8, and 6\,mHz respectively \cite{Stephan}.
\label{fig:noise}}
\end{figure}

The free torsional oscillation period of our pendulum, with
the Cu plate $\approx 1\,$cm away, is 830\,s.  The calculated
moment of inertia of our Si pendulum is $I = 135$\,g\,cm$^2$.
Four masses rotate around the vacuum chamber act
as a gravitational calibration source for converting
angular deflections into torques \cite{Stephan}.
Figure~\ref{fig:noise} shows typical torque noise data
for our pendulum when under electrostatic feedback.
Also shown is the instrumental limit given by the sum
of the pendulum thermal noise, with a quality factor $Q = 4000$, 
and the autocollimator noise converted into torque.
The LISA-equivalent requirement is the goal acceleration 
noise sensitivity \cite{srd4} 
converted into torque assuming a proof mass of $M = 1.96$\,kg
and a moment arm $l = 33\,$mm.

Variations in the electronic work function between the Au coated
Si pendulum and Cu plate can be exaggerated in our apparatus by decreasing
the separation between them.  
We measure the spatial average surface potential 
between our pendulum and one half
of the Cu plate by finding the minimum of the voltage-torque curve
given by the following expression:
\begin{equation}\label{eqn:torque}
N = \frac{1}{2}\frac{d C}{d\theta} \left( V - V_{SP} \right)^2,
\end{equation}
\noindent where the capacitance is given by
$C (\theta) \approx \epsilon_0 A / s$, with
$A = 2.18 \times 10^{-3}\,\rm{m}^2$ 
the half-plate-pendulum area overlap, $s$ is the
plate-pendulum separation, $V$ is the applied electric potential
on the Cu plate (with the pendulum at instrumental ground), and $V_{SP}$
is the derived surface potential difference.  In practice, we measure
the torque via the applied voltage on the control electrode required
to keep the pendulum parallel to the Cu plate.
By changing the voltage on the Cu plate we map out a parabola and
fit for the surface potential value.
We have consistently measured $\approx 120\,$mV on the left half of
the Cu plate and $\approx 25\,$mV on the right half.  
The value we determine is the sum of all contact potential junctions
as well as the physical surface potential difference between the two
Au coated surfaces.  Therefore our results represent an upper limit
on the physical surface potential fluctuations between the two surfaces.
The electrical paths to our DAC of both halves of the Cu plate are similar.  
The connections external to the vacuum chamber
were verified of not causing this discrepancy by swapping them.


\begin{figure}
  \includegraphics[width=0.75\columnwidth,angle=90]{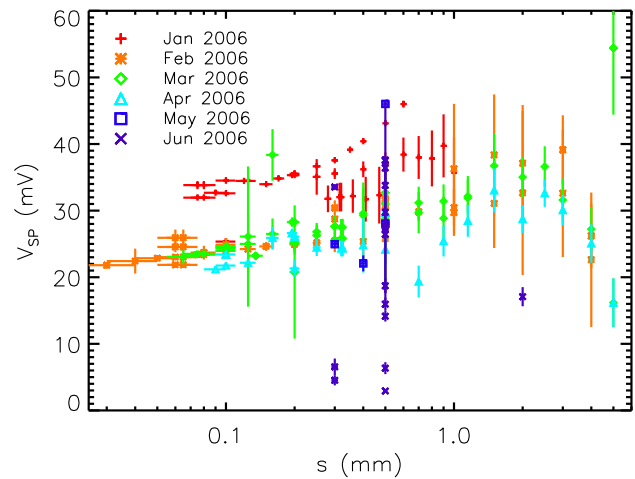}
\caption{
(color online) Measurements of the surface potential between the pendulum
and the right half of the Cu plate as a function of the plate-pendulum separation.
Both halves of the Cu plate have similar separation and temporal characteristics.
Variation with separation may be explained by spatial variations
in the surface potential.
\label{fig:distance}}
\end{figure}

We have noticed that the derived surface potential appears to vary somewhat with 
the plate-pendulum separation.  Figure~\ref{fig:distance} contains
surface potential determinations against $s$ over the course
of the first half of the year 2006.  We suspect that this effect
is due to spatial variation of the patch fields
across the Cu plate and pendulum.
As the plate-pendulum separation is changed, different patches
are averaged over resulting in slight differences in the surface
potential value.
Variations such as these have been observed with
Kelvin probe measurements which determine patch fields on the order
of 1\,mm in size for similar materials \cite{Norna}.  
The slow variation on the timescale of weeks indicates for LISA
that the surface potentials will need to be periodically measured
and corrected for by applying small DC biases on the control electrodes 
\cite{LISA_Tech_Status}.  
In our apparatus we apply DC biases to each half of the Cu 
plate to cancel the surface potential.  When the appropriate
potentials are not applied the torque noise measured from
the pendulum is increased (see figure~\ref{fig:noise})
we believe due to the amplification of voltage fluctuations
via the variation of Equation~\ref{eqn:torque}
with $V = 0$ rather than $V \approx V_{SP}$:
\begin{equation}
\delta N = \left| \frac{d C}{d\theta} 
	(V - V_{SP}) \right| \sqrt{\delta V^2 + \delta V_{SP}^2}.
\end{equation}

\begin{figure}
  \includegraphics[width=0.75\columnwidth,angle=90]{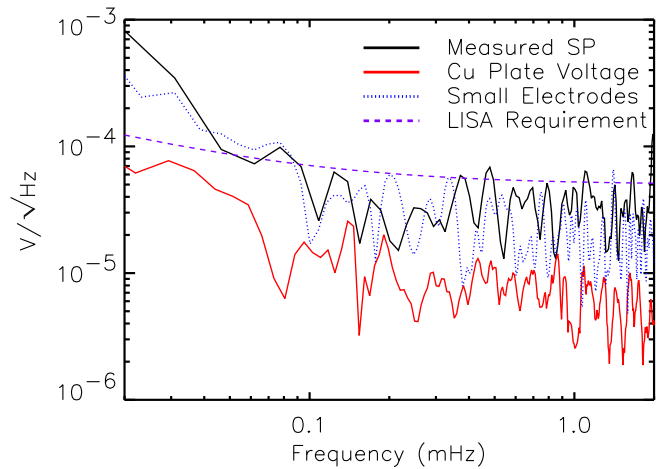}
\caption{
(color online) Measured surface potential fluctuations (dark solid) using the
method described in the text have a level of $30\,\uVhz$ rising
as $1/f$ below 0.1\,mHz.  
The LISA voltage fluctuation requirement (dashed) is $50\,\uVhz$ 
rising as $1/\sqrt{f}$ below 0.1\,mHz \cite{LISA_Tech_Status}.
The red (light solid) trace is the voltage
noise on the split Cu plate measured electronically.
Using the small control electrodes (blue, dotted) for control does not
significantly reduce the measured noise level even though
the contribution due to output voltage noise has been reduced by a factor 
$\sim 3$.
\label{fig:sp}}
\end{figure}

A faster determination of
the surface potential comes by realizing that only two applied voltages
need to be used.  
These two voltages are chosen to be symmetric about the surface
potential and yield the same torque on the pendulum, 
e.g., $V_\pm = V_0 \pm V_a$, where $V_0$ is a 
guess for the surface potential value from previous measurements, 
and $V_a \approx 0.3\,$V is chosen to yield a suitable torque on the pendulum.
We switch between these two potentials and record the torque on the
pendulum, which remains relatively constant by construction.
If the surface potential changes with time, then also will the measured torque
difference when switching between the two voltages.
Therefore the torque difference is a measure of the surface potential.
Mathematically the relation is
\begin{equation}\label{eqn:sp}
V_{SP} = V_0 - \frac{(N_+ - N_-) } {2 \frac{d C}{d\theta} V_a},
\end{equation}
where $N_\pm$ is the measured torque when $V_\pm$ is applied.
$d C/d\theta$ is determined
before and after each measurement and is on the order of $1000\,$pNm/V$^2$.
Figure~\ref{fig:sp} contains a measurement of 
surface potential fluctuations using this technique.  
The spectrum is white at $\approx 30\,\uVhz$
for frequencies above about 0.1\,mHz, and rises $\sim 1/f$ below.
The voltage on the Cu plate is monitored by our ADC during these measurements.
The electrically measured voltage noise on the Cu plate shows a similar
spectral shape, with a level $\approx 10\,\uVhz$.
The current LISA requirement for voltage fluctuations is $50\,\uVhz$
above 0.1\,mHz, rising slowly at lower frequencies~\cite{LISA_Tech_Status}.
Our upper limit on the surface potential fluctuations
meets the LISA requirement at frequencies above 0.1\,mHz, with some
excess at low frequencies.

The noise in the voltage applied to the Cu plate 
contributes to the measured surface potential value in Figure~\ref{fig:sp}.  
However, the electrically measured value of the voltage noise ($\approx 10\,\uVhz$) 
is below that of the measured surface potential
which implies that this is not the dominant source of noise.
Using a slightly noisier source ($\approx 20\,\uVhz$) 
does not increase the measured surface potential fluctuations.

The voltage noise on the control electrode will 
also 
contribute to the measured surface potential fluctuations.
By having a smaller area overlap with the pendulum, the smaller control electrodes
provide a mechanism to reduce the effect of any control voltage noise.
Given the difference in areas and lever armlengths of the
smaller control electrodes, the noise level contribution
from the smaller control electrodes should be a factor of $\sim 3$ less than that from
the larger control electrodes.  As shown in Figure~\ref{fig:sp} there does
not appear to be any substantial change in the surface potential
level when operating with the smaller control electrodes. 



\begin{figure}
  \includegraphics[width=0.75\columnwidth,angle=90]{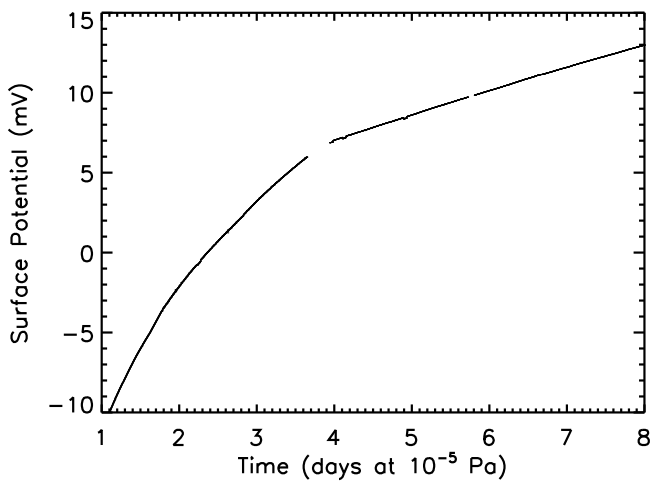}
\caption{
Surface potential measurements after venting to nitrogen, atmosphere,
and pumping back to $\approx 10^{-5}$\,Pa, with a mild bake-out at $50^\circ$.
An exponential fit to this data gives a time constant of about 2.5\,days.
The drift rate after 30\,days at this pressure was measured to be $\approx 0.30\,$mV/day
and after 50\,days it was $\approx 0.15\,$mV/day.
This slow drift of the surface potential is likely due to contamination
located on our Au coated surfaces.
\label{fig:drift}}
\end{figure}

The white nature of the fluctuations is unusual, 
and should fall off at higher frequencies.  
Our measurements are limited in frequency by the response
of our feedback loop when switching between $V_{+}$ and $V_{-}$.
It is quite possible that by using a modified Kelvin probe, 
or a Kelvin probe style torsion pendulum, one can improve the frequency span
we have presented here.

An estimation of the measurement sensitivity of our procedure is obtained
by applying a constant voltage, i.e., $V_a = 0$, to the Cu plate and analyzing
the data as described above (assuming $V_a = 0.1\,$V).
In this manner the fluctuations in the measured torque due to the surface potential
are suppressed since $V_0 \approx V_{SP}$.  We determine a measurement
sensitivity 
level of $6\,\uVhz$ which does not rise at frequencies below 0.1\,mHz.
This is consistent with the result from Equation~\ref{eqn:sp} using the noise level 
measured in Figure~\ref{fig:noise} at our switching frequency of 3\,mHz with a
switching amplitude of $V_a = 0.1\,$V.  
This indicates that our measurement process is not limited at $30\,\uVhz$.
It is possible that dielectric losses in the Cu-Si capacitor introduce noise
at this level.

We have reason to believe that contamination, which may lead to dielectric losses, 
is a principal contributor to the measured surface potential fluctuations.  
After venting our system to nitrogen, then atmosphere, and then
pumping back to $\approx 10^{-5}$\,Pa, and a mild bake-out at $50^\circ$,
we monitored the surface potential value for several days,
shown in Figure~\ref{fig:drift}.
It is likely that adsorbed materials are outgassing from the surfaces,
causing the surface potential value between the two surfaces to drift.
In a previous bake, while under vacuum, we noticed a change in
the surface potential of $\approx 80\,$mV before and after
baking at $\sim 50^\circ$.
A recent study \cite{prd_thermal} has shown that force noise due to the
outgassing of particles should be a small effect in LISA.  However,
outgassing related electrical effects, 
e.g., conveyed by thermal recrystallization or thermoelectricity, were not studied.

In Figure~\ref{fig:drift}
we observe fast jumps in the surface potential
as well as what appears to be a kink in the curve just before 4 days.  
The kink occurs after a period
of time in which we took calibration data, 
e.g., determining $d C/d\theta$ and $s$.
There also appear to be fast jumps in the data, 
some as large as 0.1\,mV, for which we
are uncertain of the source.
We are now in the process of taking data with an electrically isolated
pendulum and see similar steps in the deduced electrical charge on the pendulum,
quite possibly occurring due to cosmic ray showers \cite{Mitrofanov}.



We have built a torsion balance to measure small forces between closely spaced surfaces. 
We have used this instrument to measure surface potential fluctuations between 
two gold coated surfaces.  
In our setup, we find an upper limit of $\approx 30\uVhz$ 
for these fluctuations at frequencies above 0.1\,mHz. 
This result is relevant to the design of LISA and advanced LIGO.


\begin{acknowledgments}
We thank the members of E\"{o}t-Wash
and the Center for Experimental Nuclear Physics and Astrophysics
at the University of Washington
for infrastructure.
This work has been performed under contracts 
NAS5-03075 through GSFC,
1275177 through JPL,
and through NASA Beyond Einstein grant NNG05GF74G.
\end{acknowledgments}

\bibliography{surface}

\end{document}